\documentclass[preprint,round,authoryear]{elsarticle}

\usepackage{latexsym}
\usepackage{amssymb}
\usepackage{amsmath}
\usepackage[figuresright]{rotating}
\usepackage{tabularx}
\usepackage{multirow}
\usepackage{graphicx}
\usepackage{pstricks,pst-node,pst-tree}
\usepackage{algorithmic,algorithm}
\numberwithin{algorithm}{section}
\usepackage{color}

\newcolumntype{C}{>{\bfseries\centering\arraybackslash}X}
\newcolumntype{Z}{>{\centering\arraybackslash}X}

\newtheorem{Theorem}{Theorem}[section]

\newtheorem{Definition}[Theorem]{Definition}
\newtheorem{Properties}[Theorem]{Properties}

\renewcommand{\labelenumi}{(\roman{enumi})}

\begin{document}

\title{Selecting and estimating regular vine copulae and application to financial returns}

\author[tum]{J. Di\ss mann}
\author[tum]{E. C. Brechmann\corref{cor1}}
\author[tum]{C. Czado}
\author[delft]{D. Kurowicka}

\cortext[cor1]{Corresponding author. E-mail: brechmann@ma.tum.de. Phone: +49 89 289 17439. Fax: +49 89 289 17435.}

\address[tum]{Center for Mathematical Sciences, Technische Universit\"at M\"unchen,
Boltzmannstr. 3, 85747 Garching, Germany.}
\address[delft]{Department of Applied Mathematics, Delft University of Technology,
Mekelweg 4, 2628 CD Delft, Netherlands.}

\begin{abstract}
Regular vine distributions which constitute a flexible class of multivariate dependence models are discussed.
Since multivariate copulae constructed through pair-copula decompositions were introduced to the statistical community, interest in these models has been growing steadily and they are finding successful applications in various fields.
Research so far has however been concentrating on so-called canonical and D-vine copulae, which are more restrictive cases of regular vine copulae.
It is shown how to evaluate the density of arbitrary regular vine specifications.
This opens the vine copula methodology to the flexible modeling of complex dependencies even in larger dimensions.
In this regard, a new automated model selection and estimation technique based on graph theoretical considerations is presented.
This comprehensive search strategy is evaluated in a large simulation study and applied to a 16-dimensional financial data set of international equity, fixed income and commodity indices which were observed over the last decade, in particular during the recent financial crisis.
The analysis provides economically well interpretable results and interesting insights into the dependence structure among these indices.
\end{abstract}


\maketitle

\section{Introduction}

The most popular statistical dependence model is the multivariate Gaussian distribution. However there is a growing demand for non-Gaussian models especially in finance \citep{CherubiniLucianoVecchiato2004} but also in climate research (e.g., \citet{SchoelzelFriederichs2008}), environmental sciences (\citet{Salvadori2007} and \citet{Kazianka:2011:BSM:1953655.1954187}), medicine (e.g., \citet{Beaudoin2008}) and physics (e.g., \citet{SatoIchikiTakeuchi2010}) to name a few areas. With the availability of large samples of multivariate data it is possible to investigate non-Gaussian dependency models and to estimate parameters efficiently. The backbone for such models is the famous theorem by \citet{Sklar}, which allows to construct general multivariate distributions from copulae and marginal distributions. The specification of the copula can be done independently from the margins. While there is a multitude of bivariate copulae (see the books of \citet{HJoe} and \citet{IntroCopulas}), the class of multivariate copulae was quite restricted until recently. Especially two copula classes received attention, the class of elliptical copulae (\citet{FangFangKotz2002}, \citet{FrahmJunkerSzimayer2003}) and the class of Archimedean copulae \citep{Nelsen2005}. Typical elliptical copulae are the symmetric Gaussian and Student-t copulae (see for example \citet{DemartaMcNeil2005}), while the class of Archimedean copulae includes the tail-asymmetric Clayton and Gumbel copulae.

For financial applications a flexible modeling of tails is vital to assess the most common risk measure Value-at-Risk (VaR) (for a definition see  \citet{McNeilFreyEmbrechts2005}).  In particular the Gaussian copula does not allow for heavy tails and the approach suggested by \citet{Li2000} was blamed by many for contributing to the recent financial crisis (see \citet{Salmon2009}). This shows that there is a growing need for more flexible copulae. While the Student-t copula allows for symmetric tail dependence as measured by the tail dependence coefficient or tail dependence function (see for example \citet{JoeLiNikoloulopoulos2010}) it has only a single parameter to control tail dependence of all pairs of variables. Standard Archimedean multivariate copulae may be tail-asymmetric, but are governed only by a single parameter. There has been effort to extend the class of Archimedean copulae (see \citet{HJoe}, \citet{SavuTrede2010}, and \citet{Hofert2011}), however these models require additional parameter restrictions.

These problems were noted by \citet{Pair-copulaConstructions}, who started to utilize a wider class of multivariate copulae. This class is constructed using only bivariate copula specifications as dependency models for the distribution of certain pairs of variables conditional on a specified set of variables. These independent building blocks are called pair-copulae and were used to construct multivariate distributions. This approach dates back to \citet{Joe1996} and was investigated and organized systematically by \citet[2002]{Vines2}. The identification of the needed pairs of variables and their corresponding set of conditioning variables is facilitated by a sequence of trees (see for example Chapter 4 of \citet{UncertaintyAnalysis}). They called these trees regular vines (R-vines) and the corresponding multivariate distribution an R-vine distribution. For an $n$-dimensional R-vine distribution, the first tree identifies $n-1$ pairs of variables, whose distribution is modeled directly. The second tree identifies $n-2$ pairs of variables, whose distribution conditional on a single variable is modeled by a pair-copula. The conditioning variable is also determined in the second tree. The next tree again identifies pairs of variables, whose conditional distribution is specified by a pair-copula. Here the conditioning set has dimension $2$ and is also determined. Proceeding in this way the last tree determines a single pair of variables, whose distribution conditional on all remaining variables is defined by a last pair-copula. Recent developments and applications are discussed in \citet{Handbook}. \citet{Czado2010} provides a current survey about these statistical model classes and \citet{JoeLiNikoloulopoulos2010} investigate and discuss tail dependence properties of vine distributions. 

\citet{Pair-copulaConstructions} popularized two subclasses of regular vines, canonical vines (C-vines) and drawable vines (D-vines). C-vines possess star structures in their tree sequence, while D-vines have path structures. \citet{UncertaintyAnalysis} focused on vine distributions with Gaussian pair-copulae, but \citet{Pair-copulaConstructions} allowed for different pair-copula families, such as the bivariate Student-t copula, bivariate Gumbel and bivariate Clayton copula. While D-vine based models are started to be used in many applications (\citet{FischerKockSchluterWeigert2009}, \citet{BayesionInferenceForCopulas}, \citet{CholleteHeinenValdesogo2009}, \citet{HofmannCzado2010},  \citet{Mendes2010}, \citet{Salinas-Gutierrez:2010}, \citet{RePEc:cgr:cgsser:02-01}, \citet{mercier:statistical}, \citet{SmithMinCzadoAlmeida2010}), C-vines are less commonly used (\citet{HeinenValdesogo2009}, \citet{CzadoSchepsmeierMin2010}); \citet{NikoloulopoulosJoeLi2012} consider both classes.

Estimation in C- and D-vine copula models is often facilitated using maximum likelihood. Since this will require optimization with respect to at least $n(n-1)/2$ parameters, it is important to provide good starting values for the optimization. For this purpose a fast sequential estimation procedure was suggested and implemented in \citet{Pair-copulaConstructions}, whose asymptotic properties are investigated in \citet{Haff2010}. Since bootstrapping or inversion of high dimensional Hessian matrices are required to obtain interval estimates, Bayesian approaches have been followed for parameter estimation \citep{BayesionInferenceForCopulas} and pair-copula selection in specified D-vine copula models (\citet{MinCzado2009} and \citet{SmithMinCzadoAlmeida2010}). 

However the class of R-vine distributions is much larger than the class of D- and C-vine distributions and currently there are very few applications of R-vines. One reason for this is the enormous number of possible R-vine tree sequences (see \citet{NumberVines}) to choose from. The importance of a good selection choice has also been noted by \citet{RePEc:cir:cirwor:2009s-21}. This provides the starting point of this paper. We develop an automated strategy of jointly searching for an appropriate R-vine tree structure, the pair-copula families and the parameter values of the chosen pair-copula families. It is a sequential approach starting by identifying the first tree, its pair-copula families and estimating their parameters. Based on this the specification of the second tree utilizes transformed variables. The applied transformations depend on the choices made in the first tree. In this manner all trees together with their choice of pair-copula families and corresponding parameters are made. For each tree selection we use a maximum spanning tree algorithm, where edge weights are chosen appropriately to reflect large dependencies. Pair-copulae are chosen independently. Here we use the Akaike information criterion \citep{Akaike}, which performs well in this context (see \citet[Chapter 5]{Brechmann2010}). Finally the corresponding pair-copula parameter estimation follows the same sequential estimation approach as suggested for D- and C-vine copula distributions in \citet{Pair-copulaConstructions}.

With this automated search strategy we identify for multivariate data on the $n$-dimensional cube $[0,1]^n$ useful multivariate copula models, as we show in a large simulation study and meaningful models arise for the application considered later.

Once an appropriate R-vine distribution is found for a data set we perform maximum likelihood estimation for the parameters using the sequential estimates as starting values. We also like to perform this task in an automated setup. This requires an efficient storage of the R-vine tree specification, its pair-copula families and the corresponding parameters. This is facilitated in a set of lower triangular matrices and we proof how the corresponding joint density making up the likelihood can be evaluated recursively. This setup is also used to provide an algorithm for simulating from an R-vine distribution. Pseudo code for the corresponding algorithms is given.  

Finally we like to note that the developed search strategies are able to work not only in an automated fashion but also for higher dimensional problems. Before full maximum likelihood estimation was implemented for problems in at most 10 dimension. In our 16-dimensional application to financial data we show the usefulness of our approach and demonstrate that R-vine distributions provide better fit than C- and D-vines for this data set. These results have already spawned new research on finding more parsimonious specifications, which replace higher pair-copulae by independence copulae. See \citet{BrechmannCzadoAas2010} for details. This allows us to extend the implementation to higher dimensions, which are especially needed for the risk assessment of larger financial portfolios.

To summarize, our contributions: We develop novel algorithms for evaluating an R-vine density and simulating from specified R-vines. That is we effectively provide statistical inference techniques for R-vines. We further propose an innovative R-vine selection and estimation method and thus, for the first time, allow to actually \textit{select} and \textit{fit} arbitrary non-Gaussian R-vines to data.
This is exploited to analyze the returns of important financial indices.

The paper is organized as follows: Section \ref{sect:rvines} introduces R-vine distributions and copulae. Necessary background from graph theory can be found in \citet{Diestel2006}. Then the efficient storage of the R-vine specification and its statistical inference are developed. Selection of the R-vine tree structure, the  pair-copula families and its parameters are tackled in Section \ref{sect:rvineselect}. This includes a simulation study presented in \ref{app:simstudy} and shows that the proposed models by the search strategy are reasonable. The search and estimation algorithm is then successfully applied to a 16-dimensional financial data set involving daily equity, fixed income and commodity indices. In addition to sequential estimates full ML estimates are also provided. The paper closes with a summary and discussion.

\section{Parametric regular-vine distributions}\label{sect:rvines}

\subsection{Regular vines}

We begin this section with the theoretical background of a \textit{regular vine (R-vine)}, we then give its representation as a matrix and show how the R-vine copula density can be written in a convenient way using this matrix form. The following summarizes some definitions and results from \citet{Vines2}, \citet[Part 4]{Vines} and \citet[Chapter 4.4]{UncertaintyAnalysis}, where a tree is a graph in which each two nodes are connected by a unique sequence of edges.

\begin{Definition}[R-vine]
	\label{Def:Vine}
	$\mathcal{V} = (T_1, \ldots, T_{n-1})$ is an R-vine on $n$ elements if
	\begin{enumerate}
		\item $T_1$ is a tree with nodes $N_1 = \{1, \ldots, n\}$ and a set of edges denoted $E_1$.
		\item For $i=2, \ldots, n-1$, $T_i$ is a tree with nodes $N_i = E_{i-1}$ and edge set $E_i$.
        \item For $i = 2, \ldots, n-1$ and $\{a,b\} \in E_i$ with $a = \{ a_1, a_2\}$ and $b = \{ b_1, b_2\}$
		it must hold that $ \#(a \cap b) = 1$ \emph{(proximity condition)}.
	\end{enumerate}
\end{Definition}

In other words, an R-vine on $n$ elements is a nested set of $n-1$ trees such that the edges of tree $j$ become the nodes of tree $j+1$. The proximity condition insures that two nodes in tree $j+1$ are only connected by an edge if these nodes share a common node in tree $j$. We notice that the set of nodes in the first tree contains all indices $1,...,n$, while the set of edges is a set of $n-1$ pairs of these indices. In the second tree the set of nodes contains sets of pairs of indices and the set of edges is built of pairs of pairs of indices, etc.

To further study properties of R-vines we define three sets associated with its edges. The \textit{complete union} of an edge is a set of all indices that this edge contains. If two nodes $a$ and $b$ are joined by an edge, then the \textit{conditioned} and \textit{conditioning sets} of this edge are the symmetric difference and the intersection of the complete unions of $a$ and $b$, respectively. 

\begin{Definition}[Complete union, conditioning and conditioned sets of an edge]\label{def:ConSet}
	The complete union of an edge $e_i \in E_i$ is the set $U_{e_i}=
\{ n \in N_1 |  \exists e_{j} \in E_j, j = 1, \ldots, i-1, \text{with } n \in e_{1} \in e_{2} \in \ldots \in e_{i-1} \in e_i \}
\subset N_1$.
\mbox{For} $e_i =$ $ \{a,b\} \in E_i$, $a,b \in E_{i-1}$,  $ i = 1, \ldots , n-1$, the conditioning set of an edge $e_i$ is $D_{e_i} = U_a \cap U_b$,
	and the conditioned sets of an edge $e_i$ are $C_{e_i,a} = U_a \setminus D_{e_i}$,
		$C_{e_i,b} = U_b \setminus D_{e_i}$ and
		$C_{e_i}    = C_{e_i,a} \cup C_{e_i,b} = U_a \triangle U_b$,
	where $A \triangle B := (A \setminus B) \cup (B \setminus A)$ denotes the symmetric difference of two sets.
\end{Definition}

\begin{figure}[t]
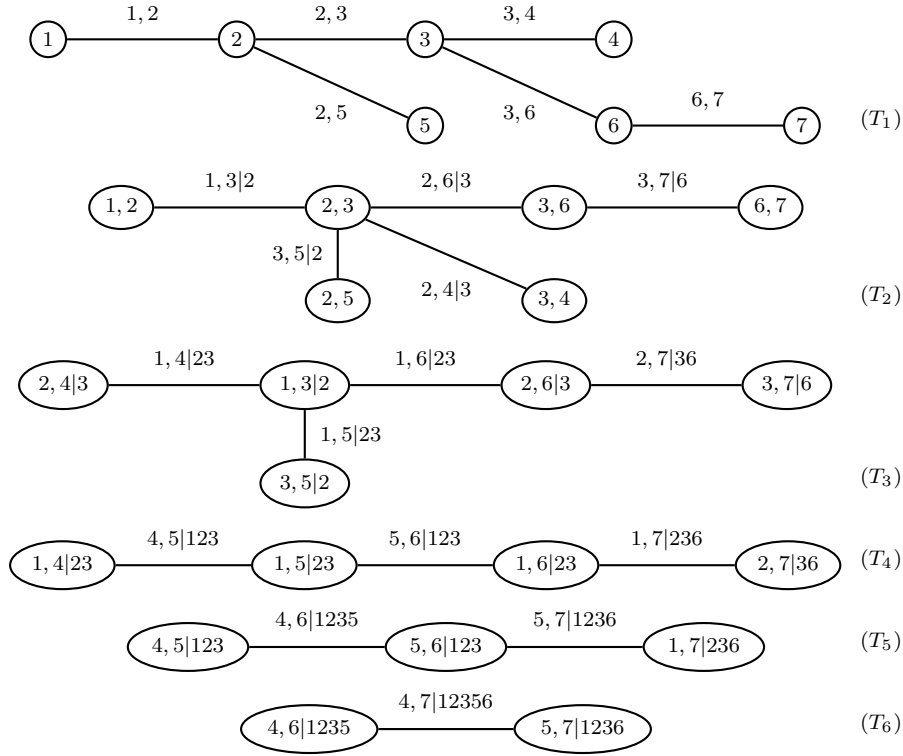

\footnotesize
\begin{align}
\psmatrix[colsep=2cm,rowsep=0.5cm,mnode= circle]
1&2&3&4\\
&&5&6&7
\ncline{1,1}{1,2}^{1,2}
\ncline{1,2}{1,3}^{2,3}
\ncline{1,3}{1,4}^{3,4}
\ncline{1,2}{2,3}_{2,5}
\ncline{1,3}{2,4}_{3,6}
\ncline{2,4}{2,5}^{6,7}
\endpsmatrix
\tag{$T_1$}
\end{align}

\begin{align}
\psmatrix[colsep=2cm,rowsep=0.5cm,mnode= oval]
1,2&2,3&3,6&6,7\\
&2,5&3,4
\ncline{1,1}{1,2}^{1,3|2}
\ncline{1,2}{1,3}^{2,6|3}
\ncline{1,3}{1,4}^{3,7|6}
\ncline{1,2}{2,3}_{2,4|3}
\ncline{1,2}{2,2}\tlput{3,5|2}
\endpsmatrix
\tag{$T_2$}
\end{align}

\begin{align}
\psmatrix[colsep=2cm,rowsep=0.5cm,mnode= oval]
{2,4|3}&{1,3|2}&{2,6|3}&{3,7|6}\\
&{3,5|2}
\ncline{1,2}{1,3}^{1,6 | 23}
\ncline{1,3}{1,4}^{2,7 | 36}
\ncline{1,2}{2,2}\trput{1,5 | 23}
\ncline{1,1}{1,2}^{1,4|23}
\endpsmatrix
\tag{$T_3$}
\end{align}

\begin{align}
\psmatrix[colsep=1.8cm,rowsep=1.5cm,mnode= oval]
{1,4|23} &{1,5 | 23}&{1,6 | 23}&{2,7 | 36}
\ncline{1,2}{1,3}^{5,6|123}
\ncline{1,1}{1,2}^{4,5|123}
\ncline{1,3}{1,4}^{1,7|236}
\endpsmatrix
\tag{$T_4$}
\end{align}

\begin{align}
\psmatrix[colsep=1.8cm,rowsep=1.5cm,mnode=oval]
{4,5|123}&{5,6|123}&{1,7|236}
\ncline{1,1}{1,2}^{4,6|1235}
\ncline{1,2}{1,3}^{5,7|1236}
\endpsmatrix
\tag{$T_5$}
\end{align}

\begin{align}
\psmatrix[colsep=1.8cm,rowsep=1.5cm,mnode=oval]
{4,6|1235}&{5,7|1236}
\ncline{1,1}{1,2}^{4,7|12356}
\endpsmatrix
\tag{$T_6$}
\end{align}

\caption{An example R-vine on seven variables. At each edge $e = \{a,b\} \in E_i$, the terms $C_{e,a}$ and $C_{e,b}$ are separated by a comma and given to the left of the `$|$' sign, while $D_e$ appears on the right.}
\label{fig:RVine}
\end{figure}

The complete union of the edge $a$ between $(1,2)$ and $(2,3)$ in tree $T_2$ shown in Figure \ref{fig:RVine} is $\{1,2,3\}$, since for instance $1\in\{1,2\}\in\{\{1,2\},\{2,3\}\}=\{a,b\}$ and $3\in\{2,3\}\in\{\{1,2\},\{2,3\}\}=\{a,b\}$, and the complete union of the edge $b$ between $(2,3)$ and $(3,6)$ is $\{2,3,6\}$.
The conditioning and the conditioned sets of the edge joining $a$ and $b$ are $\{2,3\}$ and $\{1,6\}$, respectively.

The conditioned and conditioning sets of all edges of $\mathcal{V}$ are collected in a set called \textit{constraint set}. Each element of this set is composed of a pair of indices corresponding to the conditioned set and a set containing indices corresponding to the conditioning set.   

\begin{Definition}[Constraint set]\label{def:ConSets}
The constraint set for $\mathcal{V}$ is a set:
\[
\mathcal{CV} = \big \{ ( \{C_{e,a}, C_{e,b} \} , D_e) |  e \in E_i, e = \{a,b\}, i = 1, \ldots, n-1 \big \}.
\]
\end{Definition}

It is convenient to enumerate nodes of the trees in an R-vine using their conditioned and conditioning sets. In Figure \ref{fig:RVine} each edge of the R-vine has been assigned with its conditioned sets printed before `$|$' and the conditioning set shown after `$|$'. Moreover we notice that the constraint set of an R-vine $\mathcal{CV}$ contains all necessary information needed to distinguish it from other R-vines.

Two special types of R-vines namely the \textit{canonical (C-)} and the \textit{D-vine} have been used extensively in the literature. A D-vine is an R-vine for which the first tree has nodes with degree two or less (path structure). A C-vine is an R-vine which contains a node with maximal degree in each tree (star structure). It is convenient to work with these two R-vine types as the first tree (D-vine) and the ordering of the root nodes (C-vine) determine their structure completely.

R-vines have many interesting properties that can be found in \citet{Vines} and \citet{Handbook}.

%\begin{Properties}
%\begin{enumerate}
%\item Every pair of indices appears exactly \textit{once} as a conditioned set of an edge in an R-vine.
%\item If the conditioned sets of the two edges $a$ and $b$ in an R-vine are equal, then $a=b$.\label{lemma:samecondset}
%\end{enumerate}
%\end{Properties}

\subsection{Regular vine copulae}

The graphical structure of R-vines is used to specify necessary copulae for a so-called pair-copula construction, where a copula is a multivariate distribution on the unit hypercube $[0,1]^n$ with uniform marginal distributions (see \citet{HJoe} and \citet{IntroCopulas}).
To build an R-vine copula one must specify $n-1$ unconditional bivariate copulae between variables indexed by the conditioned sets of the edges in the first tree of the R-vine. For the second tree of the R-vine one needs to specify the bivariate copulae between variables indexed by the conditioned sets conditional on variables indexed by the conditioning sets of edges of R-vine. We formally define the R-vine copula specification corresponding to an R-vine as in \citet{Vines}.

\begin{Definition}[R-vine copula specification]\label{def:RVineSpecification}
$(\boldsymbol{F}, \mathcal{V}, B)$ is an R-vine copula specification if $\boldsymbol{F} = (F_1, \ldots, F_n)$ is a vector of continuous invertible distribution functions, $\mathcal{V}$ is an $n$-dimensional R-vine and
$B = \{ B_e |  i = 1, \ldots, n-1;  e \in E_i \}$ is a set of copulae with $B_e$ being a bivariate copula, a so-called pair-copula.
\end{Definition}

A joint distribution $F$ of a random vector $(X_1, \ldots, X_n)$ is said to realize an R-vine copula specification $(\boldsymbol{F}, \mathcal{V}, B)$ or exhibit R-vine dependence if, for each $e \in E_i$, $i = 1, \ldots, n-1$, $e = \{a,b\}$, $B_e$ is the bivariate copula of $X_{C_{e,a}}$ and $X_{C_{e,b}}$ given $\boldsymbol{X}_{D_e} = \{ X_i | i \in D_e\} $, where it is assumed that this conditional copula is independent of the conditioning variables $\boldsymbol{X}_{D_e}$ (see \citet{Pair-copulaConstructions} and \citet{HaffAasFrigessi2010}). We call such a distribution also an \textit{R-vine distribution}. Additionally, the marginal distribution of $X_j$ has to be $F_j$ for $j = 1, \ldots, n$. We denote the copula density of the copula $B_e$ for the edge $e=\{ a,b\}$ as $c_{C_{e,a}, C_{e,b} | D_e}$.

For the R-vine from Figure \ref{fig:RVine} we need to assign six unconditional copulae $c_{1,2}, c_{2,3}$, $c_{3,4}, c_{2,5}, c_{3,6}$ and $c_{6,7}$ in the first tree, five conditional copulae in the second tree $c_{1,3|2}, c_{2,6|3}$, $c_{3,7|6}, c_{3,5|2}$ and $c_{2,4|3}$, etc. All copulae can be of a different type and their parameters can be specified independently from each other. However, since the copulae specified in a tree will affect the conditioned variables used in later trees the choice of the different copulae will influence each other.

The density of an R-vine copula specified through assigning appropriate bivariate copulae to edges of the R-vine has been shown in \citet[2002]{Vines2} to be equal to the product of conditional and unconditional copulae assigned to its edges.

\begin{Theorem}\label{theorem:LikelihoodRVine}
	Let $(\boldsymbol{F}, \mathcal{V}, B)$ be an R-vine copula specification on $n$ elements. There is a unique distribution $F$ that realizes this R-vine copula specification with density
	\begin{equation}
	\begin{split}
		\label{eq:RVineDensity}
		& f_{1 \ldots n}(\boldsymbol{x}) =  \\
		& \prod_{k=1}^n f_k(x_k) \prod _{i = 1} ^{n-1} \prod _{e \in E_i} c_{C_{e,a}, C_{e,b} | D_e} \big( F_{C_{e,a} | D_e}(x_{C_e,a} | \boldsymbol{x}_{D_e}), F_{C_{e,b} | D_e}(x_{C_e,b} | \boldsymbol{x}_{D_e}) \big),
	\end{split}
	\end{equation}
	where $\boldsymbol{x}=(x_1, \ldots, x_n)$, $e=\{a,b\}$ and $\boldsymbol{x}_{D_e}$ stands for the variables in  ${D_e}$, i.e., $ \boldsymbol{x}_{D_e} = \{ x_i | i \in D_e \}$. Moreover $f_i$ denotes the density of $F_i$ for $i=1,\ldots,n$.
\end{Theorem}

Notice that the copulae in \eqref{eq:RVineDensity} are indexed by elements of the set $\mathcal{CV}$ (see Definition \ref{def:ConSets}). To obtain the conditional distributions $F_{C_{e,a} | D_e}(x_{C_{e,a}} | \boldsymbol{x}_{D_e})$ and $F_{C_{e,b} | D_e}(x_{C_{e,b}} | \boldsymbol{x}_{D_e})$ let $E_i\ni e =\{a,b\}, a=\{a_1, a_2\}, b= \{b_2,b_2\}$ be the edge which connects $C_{e,a}$ with $C_{e,b}$ given the variables $D_e$. \citet{Joe1996} showed that
\begin{equation}
\begin{split}
	F_{C_{e,a} | D_e}(x_{C_{e,a}} | \boldsymbol{x}_{D_e}) &=
		\frac{\partial C_{C_{a} | D_a} 
		(F_{C_{a,a_1} | D_a}(x_{C_{a,a_1}} | \boldsymbol{x}_{D_a}),
		F_{C_{a,a_2} | D_a}(x_{C_{a,a_2}} | \boldsymbol{x}_{D_a})) }
		{\partial F_{C_{a,a_2} | D_a}(x_{C_{a,a_2}} | \boldsymbol{x}_{D_a})}\\
		&=:
		h(F_{C_{a,a_1} | D_a}(x_{C_{a,a_1}} | \boldsymbol{x}_{D_a}),
		F_{C_{a,a_2} | D_a}(x_{C_{a,a_2}} | \boldsymbol{x}_{D_a})),
\end{split}
\label{eq:hfunc}
\end{equation}
where $F_{C_{a,a_1} | D_a}(x_{C_{a,a_1}} | \boldsymbol{x}_{D_a})$ and $F_{C_{a,a_2} | D_a}(x_{C_{a,a_2}} | \boldsymbol{x}_{D_a}))$ have to be obtained recursively as shown in the next section. The notation of the $h$-function is introduced for convenience.

Similarly, we obtain $F_{C_{e,b} | D_e}(x_{C_{e,b}} | \boldsymbol{x}_{D_e})$. We call $F_{C_{e,a} | D_e}(x_{C_{e,a}} | \boldsymbol{x}_{D_e})$ and $F_{C_{e,b} | D_e}(x_{C_{e,b}} | \boldsymbol{x}_{D_e})$ \textit{transformed variables}.

For C- and D-vines the density \eqref{eq:RVineDensity} can be rewritten in a more convenient way. For more information on how to exploit the structure of C- and D-vines see \citet{BergAas2007}, \citet[2011]{BayesionInferenceForCopulas}\nocite{MinCzado2009} and \citet{CzadoSchepsmeierMin2010}.

\subsection{Matrix representation of regular vines}

To develop statistical inference algorithms for R-vines we need a convenient way of representing an R-vine. Storing the nested set of trees is too expensive and does not allow for an easy way to describe inference algorithms.

\citet{Morales} uses a lower triangular matrix to store an R-vine. The idea is to store the constraint set of an R-vine in columns of an $n$-dimensional lower triangular matrix. We hence specify how the information from the lower triangular matrix should be read by defining a constraint set for the matrix.
In the next section we introduce a way how the structure of R-vine matrices can be used to encode corresponding pair-copula types and parameters.
While \citet{Morales} used the matrix representation of R-vines for counting the number of different R-vines, we will subsequently exploit this structure for likelihood computation and a sampling procedure.

\begin{Definition}[Matrix constraint set]
	Let  $M =(m_{i,j})_{i,j=1,\ldots n}$ be a lower triangular matrix.
	The $i$-th constraint set for $M$ is
	\begin{equation}
		\mathcal{C}_M (i) = \big\{ (\{ m_{i,i}, m_{k,i} \}, D ) | k = i+1, \ldots, n, D= \{ m_{k+1,i}, \ldots, m_{n,i} \} \big\}
		\label{eq:matrixconstrset}
	\end{equation}
	for $i = 1, \ldots, n-1$.
	If $k=n$ we set $D = \emptyset$.
	The constraint set for matrix $M$ is the union $\mathcal{C} M = \mathcal{C}_M(1) \cup \ldots \cup \mathcal{C}_M(n-1)$.
	For the elements of the constraint set $(\{ m_{i,i}, m_{k,i} \}, D ) \in \mathcal{C} M$ we call $\{ m_{i,i}, m_{k,i} \}$ the conditioned set and $D$ the conditioning set.
\end{Definition}

Every element of the constraint set is made up of an diagonal entry $m_{i,i}$, an entry in the same column below the diagonal $m_{k,i}$ and all the elements following in that column $\{ m_{k+1,i}, \ldots, m_{n,i} \}$,  $k = i+1, \ldots, n$, $i = 1, \ldots, n$.

To demonstrate this idea, we can compare the constraint sets defined by the example matrix $M^*$ with the constraint sets of the R-vine in Figure \ref{fig:RVine}.

\begin{equation}
M^* =
\left( \begin{array}{ccccccc}
7\\
4&4\\
5&6&6\\
1&5&5&5\\
2&1&1&1&1\\
3&2&2&3&3&3\\
6&3&3&2&2&2&2\\
\end{array} \right).
\label{eq:Matrix}
\end{equation}

In the first column of $M^*$ we have the diagonal entry $m_{1,1} = 7$ and the element $m_{4,1} = 1$ in the fourth row. According to the definition above this gives $(\{ 7,1\},\{2,3,6\} ) \in \mathcal{C} M^*$ which corresponds to the constraint set of the rightmost edge of $T_4$ in the R-vine in Figure \ref{fig:RVine}.

Before we formally define an R-vine matrix (that will be shown to code all information included in an R-vine) we need two sets that will help us characterize the matrix form and will ensure the proximity condition required for R-vines (see Definition \ref{Def:Vine}). For a lower triangular matrix $M =(m_{i,j})_{i,j=1,\ldots n}$ set for $i=1,...,n-1,$
\begin{align*}
	B_M(i) &:= \big\{ (m_{i,i}, D) | k = i+1, \ldots, n ; D=\{m_{k,i}, \ldots, m_{n,i} \} \big\}, \\
	\widetilde{B}_M(i) &:= \big\{ (m_{k,i}, D) | k = i+1, \ldots, n ; D= \{m_{i,i}\} \cup \{m_{k+1,i}, \ldots, m_{n,i} \} \big\}.
\end{align*}

Now we can define an R-vine matrix.

\begin{Definition}[R-vine matrix]
	\label{def:r-vine-matrix}
	A lower triangular matrix $M =(m_{i,j})_{i,j=1,\ldots n}$ is called an R-vine matrix if
for $i = 1, \ldots, n-1$ and for all $k = i+1, \ldots, n-1$ there is an $j$ in $i+1, \ldots, n-1$ with
\begin{equation}
			(m_{k,i}, \{ m_{k+1,i}, \ldots, m_{n,i} \}) \in B_M(j) \text{ or } \in \widetilde{B}_M(j).
\label{eq:rvinematrix}
\end{equation}
\end{Definition}

It can be shown that the following two properties follow from \eqref{eq:rvinematrix}:
\begin{enumerate}
		\item $\{m_{i,i}, \ldots, m_{n,i}\} \subset \{m_{j,j}, \ldots m_{n,j}\}$ for $1 \leq j < i \leq n$,
		\item $m_{i,i} \not \in \{m_{i+1,i+1}, \ldots, m_{n,i+1}\}$ for $ i = 1, \ldots, n-1$.
\end{enumerate}
Condition (i) states that every column contains all the entries that a column to the right contains, while condition (ii) assures that there is a new entry on the diagonal in every column. Condition \eqref{eq:rvinematrix} is the essential counterpart to the proximity condition in the definition of an R-vine (see Definition \ref{Def:Vine}). Note that \citet{Morales} used a different condition to ensure the proximity condition.

As an example, one may check that $M^*$ given in \eqref{eq:Matrix} fulfills condition \eqref{eq:rvinematrix} and is in fact an R-vine-matrix.

The following simple properties of an R-vine matrix can be seen directly from the definition.

\begin{Properties}
\begin{enumerate}
\item All elements in a column are different.
\item Deleting the first row and column from an $n$-dimensional R-vine matrix gives an $(n-1)$-dimensional R-vine matrix.
\end{enumerate}
\end{Properties}

We have seen that the matrix $M^*$ codes all information needed to represent the R-vine in Figure \ref{fig:RVine}. The proof that there is an equivalent R-vine-matrix with the same constraint set for every R-vine and vice versa can be found in \citet{Dissmann2010}. In the proof it is shown that the constraint set $\mathcal{CV}$ of an R-vine is in fact equal to the constraint set $\mathcal{C}M$ of a corresponding R-vine matrix $M$. Note however that the matrix corresponding to an R-vine is not unique. As a simple example consider the matrix obtained after an exchange of the elements 2 and 3 in the lower right 2 by 2 corner of $M^*$. It defines the same R-vine as $M^*$.

\subsection{Evaluation of the joint regular vine density}

We now use the matrix representation for R-vines presented in the previous section to make more visible which copulae have to be used to build a density of the R-vine distribution.
In particular, we provide a novel algorithm on how to efficiently evaluate the conditional distribution functions of an arbitrary R-vine copula.
This is a non-trivial task, since the order of the conditioning variables required is not obvious.
For this purpose we require an R-vine matrix that codes information about conditioned and conditioning variables. Let $M=(m_{i,j})_{i,j=1,\ldots ,n}$ be an R-vine matrix corresponding to the R-vine $\mathcal{V}$.

The R-vine distribution is a product of copulae indexed by $\mathcal{CV}$ which is equal to $\mathcal{C} M$ defined in \eqref{eq:matrixconstrset}. Hence the R-vine distribution density is:
 
\begin{equation}
\begin{split}
		\label{eq:DensRVine3}
		& f_{1 \ldots n} =\\
		&  \prod_{j=1}^n f_j \prod _{k =n-1} ^{1} \prod _{i=n}^{k+1} c_{m_{k,k}, m_{i,k} | m_{i+1,k},\ldots,m_{n,k}} \big( F_{m_{k,k} | m_{i+1,k},\ldots,m_{n,k}}, F_{m_{i,k} | m_{i+1,k},\ldots,m_{n,k}} \big),
\end{split}
\end{equation}
where arguments of all functions have been omitted to shorten the notation.

We now have to show how the conditional distributions which are arguments of bivariate copulae in \eqref{eq:DensRVine3} are obtained. We will show this in the algorithm below where the evaluation of the fully parametrical form of an R-vine distribution is described. For this purpose we first need to specify two additional square matrices $T=(t_{i,j})_{i,j=1,\ldots ,n}$ and $P=(p_{i,j})_{i,j=1,\ldots ,n}$ that will contain information about types and parameters of the bivariate copulae in \eqref{eq:DensRVine3}.

Since for all $j=1,\ldots, n-1$, $i=j+1,\ldots, n$ the entry $m_{i,j}$ of $M$ codes the copula of the variables indexed by $m_{j,j}$ and $m_{i,j}$ conditional on the variables indexed by $\{m_{i+1,j},\ldots,m_{n,j}\}$ we let $t_{i,j}$ describe the type of this copula (e.g., Normal, Clayton, etc.) and let $p_{i,j}$ contain parameters of this copula (note that some copulae require more than one parameter; we can store them, e.g., in additional matrices or using a multi-dimensional array instead of a matrix). An example of such a specification for $M^*$ (see \eqref{eq:Matrix}) is shown in Figure \ref{fig:CopulaTypeSpec}.

\begin{figure}[t]
\footnotesize
\centering
\begin{tabular}{b{0.22\textwidth}b{0.36\textwidth}b{0.4\textwidth}}
$M^* = $&$T^* = $&$P^* =$\\
\hline
$
\psmatrix[colsep=0.2cm,rowsep=0.15cm]
            [mnode=circle,linecolor=red]4\\
            7&5\\
            6&7&1\\
     [mnode=circle,linecolor=red,linestyle=dashed]5&6&7&[mnode=circle,linecolor=green]7\\
	   1&1&6&2&6\\
             2&3&3&3&2&2\\
             3&2&2&[mnode=circle,linecolor=green,linestyle=dashed]6&3&3&3
\ncbox[nodesep=.25cm,boxsize=.35,linearc=.2,linecolor=red]{5,1}{7,1}
\endpsmatrix
$
&
$
\psmatrix[colsep=0.2cm,rowsep=0.15cm]
            { }\\
            t_{2,1}&\\
            t_{3,1}&t_{3,2}&\\
             [mnode=oval,linecolor=red,linestyle=dashed]t_{4,1}&t_{4,2}&t_{4,3}&\\
	   t_{5,1}&t_{5,2}&t_{5,3}&t_{5,4}&\\
             t_{6,1}&t_{6,2}&t_{6,3}&t_{6,4}&t_{5,5}&\\
             t_{7,1}&t_{7,2}&t_{7,3}&[mnode=oval,linecolor=green,linestyle=dashed]t_{7,4}&t_{7,5}&t_{7,6}&
\endpsmatrix
$
&
$
\psmatrix[colsep=0.2cm,rowsep=0.15cm]
	   { }\\
            p_{2,1}&\\
            p_{3,1}&p_{3,2}&\\
            [mnode=oval,linecolor=red,linestyle=dashed]p_{4,1}&p_{4,2}&p_{4,3}&\\
	   p_{5,1}&p_{5,2}&p_{5,3}&p_{5,4}&\\
             p_{6,1}&p_{6,2}&p_{6,3}&p_{6,4}&p_{6,5}\\
             p_{7,1}&p_{7,2}&p_{7,3}&[mnode=oval,linecolor=green,linestyle=dashed]p_{7,4}&p_{7,5}&p_{7,6}
\endpsmatrix
$\\
\hline
\end{tabular}
\caption{The  copula with conditioned variables indexed by $\{4,5\}$ and conditioning variables indexed by $\{ 1,2,3 \}$, i.e., $c_{4,5|123}$, is of the type $t_{4,1}$ with parameter $p_{4,1}$. The copula $c_{7,6}$ is of the type $t_{7,4}$ and has the parameter $p_{7,4}$.}
\label{fig:CopulaTypeSpec}
\end{figure}

Next, we find a recursive algorithm to calculate the conditional distributions. For convenience we will assume that the diagonal entries of $M$ are ordered from $n$ to $1$, i.e., $m_{k,k} = n-k+1$. Note that the reordered matrix is equivalent to the original matrix which means it induces the same R-vine but with relabeled indices. The copula type and parameter matrices are unaffected by this reordering. To proceed, we introduce the maximum matrix of $M$ denoted by $\mathbb{M}$. It is $\mathbb{M} = (\mathbf{m}_{i,k})_{ i,k = 1, \ldots, n}$ with $\mathbf{m}_{i,k} = \max \{ m_{i,k}, \ldots, m_{n,k} \}$ for all $k = 1, \ldots, n$ and $i = k, \ldots, n$. In words, $\mathbf{m}_{i,k}$ is the maximum of all entries in the $k$-th column of $M$ from the bottom up to the $i$-th element. Note that $\mathbf{m}_{n,k} = m_{n,k}$ for all $k = 1, \ldots, n$, since $\mathbf{m}_{n,k}$ is the maximum over only one element and since the element on the diagonal is a new element in each column, it is $\mathbf{m}_{k,k} = m_{k,k} = n-k+1$ for all $k = 1, \ldots, n$.

Algorithm \ref{alg:LogLikelihood} shows how to compute the density for a given R-vine copula specification, where $h(\cdot, \cdot | t_{i,k}, p_{i,k})$ in Line \ref{alg:LogLikelihood:h} denotes the $h$-function \eqref{eq:hfunc} for the copula type $t_{i,k}$ with parameters $p_{i,k}$ and the matrices $V^{\text{direct}}$ and $V^{\text{indirect}}$ are introduced to store the arguments of the bivariate copulae in \eqref{eq:DensRVine3}, where their notation is due to the order of the arguments in Line 15.

\begin{algorithm}[t]
\caption{Density of an R-vine specification.}
\label{alg:LogLikelihood}
\begin{algorithmic}[1]
	\REQUIRE R-vine specification in matrix form, i.e., $M$, $T$, $P$, where $m_{k,k} = n-k+1,\ k=1,...,n$.
	\ENSURE Density of the R-vine distribution at $(x_1, \ldots x_n)$ for the given R-vine specification.
	\STATE Set $F = 1$.\label{alg:LogLikelihood:F1}
	\STATE Let $V^{\text{direct}} = (v^\text{direct}_{i,k} | i,k = 1, \ldots, n)$.
	\STATE Let $V^{\text{indirect}} = (v^\text{indirect}_{i,k} | i,k = 1, \ldots, n)$.
	\STATE Set $(v^\text{direct}_{n,1}, v^\text{direct}_{n,2}, \ldots, v^\text{direct}_{n,n})
				= (F_n(x_n), F_{n-1}(x_{n-1}), \ldots, F_1(x_1))$.
	\STATE Let $\mathbb{M} = (\mathbf{m}_{i,k} | i,k = 1, \ldots, n)$ with		$\mathbf{m}_{i,k} = \max \{ m_{i,k}, \ldots, m_{n,k} \}$
	for all $k = 1, \ldots, n$ and $i = k, \ldots, n$.
	\FOR[Iteration over the columns of $M$]{$k = n-1, \ldots, 1$}
		\FOR[Iteration over the rows of $M$]{$i = n, \ldots, k+1$}
			\STATE Set $z^{(1)}_{i,k} = v^\text{direct}_{i,k}$ \label{alg:LogLikelihood:z1}
			\IF{$\mathbf{m}_{i,k} =  m_{i,k}$} \label{alg:LogLikelihood:IF}
				\STATE Set $z^{(2)}_{i,k} = v^\text{direct}_{i,(n-\mathbf{m}_{i,k}+1)}$.
				 \label{alg:LogLikelihood:IF:1}
			\ELSE
				\STATE Set $z^{(2)}_{i,k} = v^\text{indirect}_{i,(n-\mathbf{m}_{i,k}+1)}$.
				 \label{alg:LogLikelihood:IF:2}
			\ENDIF
			\STATE Set $F = F \cdot c(z^{(1)}_{i,k},z^{(2)}_{i,k} | t_{i,k},  p_{i,k})$.
			\label{alg:LogLikelihood:LL}			
			\STATE Set $v^\text{direct}_{i-1,k} =  h(z^{(1)}_{i,k},z^{(2)}_{i,k} | t_{i,k}, p_{i,k})$ and $v^\text{indirect}_{i-1,k} = h(z^{(2)}_{i,k},z^{(1)}_{i,k} | t_{i,k},  p_{i,k})$.
			\label{alg:LogLikelihood:h}	
		\ENDFOR
	\ENDFOR
	\RETURN F
\end{algorithmic}
\end{algorithm}

The outer \textbf{for}-loop of the algorithm iterates over the columns of $M$ from right to left, starting with $n-1$. The inner \textbf{for}-loop iterates over the rows from the bottom up to one element below the diagonal entry of $M$. Therefore, Line \ref{alg:LogLikelihood:LL} of Algorithm \ref{alg:LogLikelihood} is executed once for every edge of the R-vine with the corresponding copula type and parameters.

Note that we do not need to initialize $(v^\text{indirect}_{n,1}, v^\text{indirect}_{n,2}, \ldots, v^\text{indirect}_{n,n})$ because it is $\mathbf{m}_{n,k} = m_{n,k}$ for all $k = 1, \ldots, n-1$ and hence, we always select a $v^\text{direct}$ in Line \ref{alg:LogLikelihood:IF} for $i=n$.

The crucial point in the algorithm is how the conditional distributions that are arguments of bivariate copulae in (\ref{eq:DensRVine3}) denoted as  $z^{(1)}_{i,k}$ and $z^{(2)}_{i,k}$ are selected.

Therefore, we show that $z^{(1)}_{i,k} = F_{m_{k,k} |  \{ m_{i+1,k}, \ldots, m_{n,k} \}}(x_{m_{k,k}} | x_{m_{i+1,k}}, \ldots, x_{m_{n,k}})$ and
	$z^{(2)}_{i,k} = F_{m_{i,k} |  \{ m_{i+1,k}, \ldots, m_{n,k} \}}(x_{m_{i,k}} | x_{m_{i+1,k}}, \ldots, x_{m_	{n,k}} )$
	for $k = n-1, \ldots, 1$ and $i = n, \ldots, k+1$.

We argue by induction and start with $i = n$ and $k$ arbitrary in $1, \ldots, i$.

It is $z^{(1)}_{n,k} = v^\text{direct}_{n,k} = F_{n-k+1}(x_{n-k+1}) = F_{m_{k,k}}( x_{m_{k,k}} )$, and since $\mathbf{m}_{n,k} = m_{n,k}$, it is $z^{(2)}_{n,k} = v^\text{direct}_{n,n-m_{n,k}+1} = F_{m_{n,k}}(x_{m_{n,k}} )$. Thereby, the statement is valid for $i = n$.

We assume that for all $n \geq i  > I$ for an $I > 2$, i.e., for all $k = i, \ldots, 1$ it is
\begin{eqnarray}
	\label{eq:direct:123}
	v^\text{direct}_{i-1,k} &=& F_{m_{k,k} |  \{  m_{i,k},  m_{i+1,k}, \ldots, m_{n,k} \}}(x_{m_{k,k}} | x_{m_{i,k}},x_{m_{i+1,k}},\ldots, x_{m_{n,k}} )
\end{eqnarray}
and
\begin{eqnarray}
	\label{eq:indirect:123}
	v^\text{indirect}_{i-1,k} &=& F_{m_{i,k} |  \{  m_{k,k},  m_{i+1,k}, \ldots, m_{n,k} \}}(x_{m_{i,k}} | x_{m_{k,k}},x_{m_{i+1,k}},\ldots, x_{m_{n,k}} ).
\end{eqnarray}

If we proceed with step $I$, the algorithm selects $z^{(1)}_{I,k} = v^\text{direct}_{I,k}$ in Line \ref{alg:LogLikelihood:z1}. By Equation (\ref{eq:direct:123}) it is $z^{(1)}_{I,k} = F_{m_{k,k} |  \{  m_{I+1,k}, \ldots, m_{n,k} \}}(x_{m_{k,k}} | x_{m_{I+1,k}},\ldots x_{m_{n,k}} )$ which proves that the algorithm selects the correct entry for $z^{(1)}_{I,k}$.

By Definition \ref{def:r-vine-matrix}, Property (iii) we know that there exists a $j$ in $k+1, \ldots, n-1$ with
\begin{equation}
	\label{eq:MaxMatProof:1}
	(m_{I,k} ,  \{  m_{I+1,k}, \ldots, m_{n,k} \} )  \in B_M(j) \cup \widetilde B _M(j).
\end{equation}

Let $(x,D) \in B_M(j)$, then $x$ and $D$ consist of elements of the $j$-th column of M. Thus, $\max \{ x, \max D \} = m_{j,j}$. This is also true for $(x,D) \in \widetilde B_M(j)$. If we take the maximum over all elements on the left and right side of \eqref{eq:MaxMatProof:1}, it must hold that $\mathbf{m}_{I,k} = m_{j,j}$, and since $m_{j,j} = n-j+1$  we know that $j = n - \mathbf{m}_{I,k} +1$. This explains the indexation of $v$ in Lines \ref{alg:LogLikelihood:IF:1} and \ref{alg:LogLikelihood:IF:2}.

Now we distinguish between the cases $(m_{I,k} ,  \{  m_{I+1,k}, \ldots, m_{n,k} \} )  \in B_M(j) $ and \\ $(m_{I,k} ,  \{  m_{I+1,k}, \ldots, m_{n,k} \} )  \in \widetilde B_M(j) $. For $(m_{I,k} ,  \{  m_{I+1,k}, \ldots, m_{n,k} \} )  \in B_M(j) $ it is
\begin{equation}
	\label{eq:MaxMatProof:Case1}
	(m_{I,k} ,  \{  m_{I+1,k}, \ldots, m_{n,k} \} ) =
		(m_{j,j} ,  \{  m_{I+1,j},  \ldots, m_{n,j} \} ) \in B_M(j).
\end{equation}
Hence, it follows $m_{I,k} = m_{j,j} = \mathbf{m}_{I,k}$. Thus, it is $m_{I,k} =  \mathbf{m}_{I,k}$ in Line \ref{alg:LogLikelihood:IF} of the algorithm, and the algorithm defines $z^{(2)}_{I,k} = v^\text{direct}_{I,(n-\mathbf{m}_{I,k}+1)} = v^\text{direct}_{I,j}$. Using the induction assumption (\ref{eq:direct:123}) it follows 
	\begin{equation*}
z^{(2)}_{I,k} = F_{m_{j,j} |  \{  m_{I+1,j},  m_{I+2,j}, \ldots, m_{n,j} \}}(x_{m_{j,j}} | x_{m_{I+1,j}},x_{m_{I+2,j}},\ldots x_{m_{n,j}} ),
	\end{equation*}
and by (\ref{eq:MaxMatProof:Case1})
	\begin{equation*}
		z^{(2)}_{I,k} = F_{m_{I,k} |  \{  m_{I+1,k}, \ldots, m_{n,k} \}}(x_{m_{I,k}} | x_{m_{I+1,k}},\ldots x_{m_{n,k}} ).
	\end{equation*}

The argumentation for $(m_{I,k} ,  \{  m_{I+1,k}, \ldots, m_{n,k} \} )  \in \widetilde B_M(j) $ is similar. This proves the statement.

\subsection{Inference of regular vines}\label{sect:rvineinf}

Having now established Algorithm \ref{alg:LogLikelihood} to evaluate a given R-vine copula density, the determination of the corresponding log likelihood expression $L$ is straightforward by substituting Line \ref{alg:LogLikelihood:F1} through ``$L=0$'' and Line \ref{alg:LogLikelihood:LL} through ``$L = L + \log c(z^{(1)}_{i,k},z^{(2)}_{i,k} | t_{i,k},  p_{i,k})$'', and by returning $L$ instead of $F$ in the last line. The log likelihood can then be used, for example, for maximum likelihood estimation of the pair-copula parameters.

For vines there is a second estimation procedure which is typically used in the literature, namely \textit{sequential estimation}. This method exploits the tree by tree structure of vines by separately estimating the parameter(s) of each pair-copula in the first tree, then computing the transformed variables for the second tree using $h$-functions, again separately estimating the conditional pair-copulae in the second tree, and so on. In doing so, only bivariate estimation is required and hence this method is quite fast. Moreover, the estimated parameters are typically good starting values for joint maximum likelihood estimation.

With regard to Algorithm \ref{alg:LogLikelihood}, this means that we only have to insert a new line before Line \ref{alg:LogLikelihood:LL}, where the copula parameter $p_{i,k}$ is estimated based on the observations $z^{(1)}_{i,k}$ and $z^{(2)}_{i,k}$ and for copula family $t_{i,k}$.

Furthermore, sampling from R-vine specifications can be performed using the inverse probability integral transform (see \citet{PIT}). E.g., in the bivariate case, let $C$ be the copula under consideration and let $v_1$ and $v_2$ be two independent uniform samples. Using the inverse of the $h$-function as defined in \eqref{eq:hfunc}, $\boldsymbol{u}=(u_1,u_2)^\prime$ given by
\begin{equation*}
u_1 = v_1,\quad \text{and}\quad
u_2 = h^{-1}(v_2,u_1) = F_{2|1}^{-1}(v_2|u_1),
\end{equation*}
then is a sample from the copula $C$ with uniform margins.

This idea can be generalized to R-vines and the corresponding algorithm is given in Algorithm \ref{alg:Sim}, where we again assume that entries of the R-vine matrix are ordered from $n$ to $1$, and in particular the selection of the different $z^{(1)}_{i,k}$ and $z^{(2)}_{i,k}$ is the same as in Algorithm \ref{alg:LogLikelihood}. More details on this can be found in \citet[Section 5.3]{Dissmann2010}.

\begin{algorithm}[t]
\caption{Simulation of an R-vine specification.}
\label{alg:Sim}
\begin{algorithmic}[1]
	\REQUIRE R-vine specification in matrix form, i.e., $M$, $T$, $P$, where $m_{k,k} = n-k+1,\ k=1,...,n$.
	\ENSURE Random observations $(x_1, \ldots, x_n)$ from the R-vine specification.

	\STATE Let $u_1, \ldots, u_n$ be independent uniform samples.
	\STATE Let $V^{\text{direct}} = (v^\text{direct}_{i,k} | i,k = 1, \ldots, n)$.
	\STATE Let $V^{\text{indirect}} = (v^\text{indirect}_{i,k} | i,k = 1, \ldots, n)$.
	\STATE Set $(v^\text{direct}_{n,1}, v^\text{direct}_{n,2}, \ldots, v^\text{direct}_{n,n})
				= (u_1, u_{2}, \ldots u_n)$.
				
	\STATE Let $\mathbb{M} = (\mathbf{m}_{i,k} | i,k = 1, \ldots, n)$ with
		$\mathbf{m}_{i,k} = \max \{ m_{i,k}, \ldots, m_{n,k} \}$
	for all $k = 1, \ldots n-1$ and $i = k, \ldots, n$.
	
	\STATE $x_1 = v^\text{direct}_{n,n}$
	
	\FOR[Iteration over the columns of $M$]{$k = n-1, \ldots, 1$}
	
		\FOR[Iteration over the rows of $M$]{$i = k+1, \ldots, n$} \label{alg:Sim:1}

			\IF{$\mathbf{m}_{i,k} =  m_{i,k}$}
				\STATE Set $z^{(2)}_{i,k} = v^\text{direct}_{i,(n-\mathbf{m}_{i,k}+1)}$.
			\ELSE
				\STATE Set $z^{(2)}_{i,k} = v^\text{indirect}_{i,(n-\mathbf{m}_{i,k}+1)}$.
			\ENDIF
			
			\STATE Set $v^\text{direct}_{n,k} = h^{-1} (v^\text{direct}_{n,k}, z^{(2)}_{i,k} | t_{i,k}, p_{i,k}) $

		\ENDFOR \label{alg:Sim:2}
		
		\STATE $x_{n-k+1} = v^\text{direct}_{n,k}$
		
		\FOR[Iteration over the rows of $M$]{$i = n, \ldots, k+1$}
			\STATE Set $z^{(1)}_{i,k} = v^\text{direct}_{i,k}$

			\STATE Set $v^\text{direct}_{i-1,k} =  h(z^{(1)}_{i,k},z^{(2)}_{i,k} | t_{i,k}, p_{i,k})$ and 
				$v^\text{indirect}_{i-1,k} = h(z^{(2)}_{i,k},z^{(1)}_{i,k}  | t_{i,k},  p_{i,k})$.			
		\ENDFOR
	\ENDFOR
	\RETURN $(x_1, \ldots, x_n)$

\end{algorithmic}
\end{algorithm}

\section{Selecting regular vine distributions}\label{sect:rvineselect}

Fitting an R-vine copula specification to a given dataset requires the following separate tasks:
\begin{enumerate}
	\renewcommand{\labelenumi}{(\alph{enumi})}
	\item Selection of the R-vine (structure), i.e., selecting which unconditioned and conditioned pairs to use. 
	\item Choice of a bivariate copula family for each pair selected in (a).
	\item Estimation of the corresponding parameter(s) for each copula.
\end{enumerate}
Since all three steps are needed for an R-vine copula specification, one way of finding the ``best'' model is to accomplish steps (b) and (c) for all possible R-vine constructions. Since the number of possible R-vines on $n$ variables increases very rapidly with $n$ ($n!/2 \times 2^{\binom{n-2}{2}}$ as shown in \citet{NumberVines}), this is not feasible. In addition to the fast growing number of possible R-vines, some methods to decide which bivariate copula family to use depend on the interpretation of plots, e.g., K- or Chi-Plots (see \citet{EverythingCopulaModeling}), and therefore need manual interaction. On the one hand, we do not use such methods to obtain objectivity and, on the other hand, this again is not feasible to do for every possible copula in every possible R-vine decomposition. In particular, in Section \ref{sec:applic} we will fit a model to a 16-dimensional dataset leaving 120 copulae to select. This is not practicable to do manually.

Therefore, we developed a sequential, heuristic method to select the tree structure of the R-vine. Since our proposed method for (a) depends on the copulae selected in (b) and estimated in (c), copula selection is covered in Section \ref{sec:CopulaSelection}. 
A simulation study to evaluate our approach is presented in \ref{app:simstudy}.

In Section \ref{sec:applic} we will apply the techniques.

\subsection{Sequential method to select an regular vine copula specification based on Kendall's tau}\label{sec:ModelSelection}

To select one possible R-vine for a given dataset it is necessary to decide for which pairs of variables we want to specify copulae. We proceed sequentially, starting by defining the first tree $T_1 = (N_1, E_1)$ for the R-vine, continuing with the second tree, and so on. The trees are selected in such a way that the chosen pairs model the strongest pairwise dependencies present (more details below). Later, we will refer to this method as the \emph{sequential method}. Since we examine every tree separately, it is not guaranteed to find a global optimum, where global optimum is meant in terms of model fit, e.g., higher likelihood, smaller AIC/BIC or superior in terms of the likelihood-ratio based test for comparing non-nested models proposed by \citet{Vuong}.
However, we think this sequential approach is reasonable because
\begin{itemize}

\item the copula families specified in the first tree of the R-vine often have the greatest influence on the model fit.
	
\item it is more important to model the dependence structure between random variables that have high dependencies correctly, because most copula families can model independence and the copulae distribution functions for parameters close to independence are very similar.

\item this approach minimizes the influence of rounding errors in later trees, which pairs with strong pairwise dependence are most prone to, e.g., when assessing the joint tail behavior of two variables. For pairs of variables close to independence, such issues are less relevant.
	
\item for real applications it is natural to assume that randomness is driven by the dependence of only some variables and not all. Therefore, if you choose the copulae with high dependence in the first trees, the transformed variables  for the later trees will often be rather independent. We exemplify this using the multivariate normal distribution, since we can easily compute conditional dependence for multivariate normal distributions using well known properties of the normal distribution (see, e.g., \citet{MVN}).

For example consider the following three jointly normal distributed random variables.
		$$
			\left(
			\begin{array}{c}
			 X_1 \\
			 X_2 \\
			 X_2
			\end{array} \right ) \sim N \left( 
			\left(
			\begin{array}{c}
			 0 \\
			 0 \\
			 0
			\end{array} \right ) ,
			\left(
			\begin{array}{ccc}
			 1 & \rho_{1,2} & \rho_{1,3} \\
			 \rho_{1,2} & 1 & \rho_{2,3}  \\
			 \rho_{1,3} & \rho_{2,3} & 1
			\end{array} \right )
			\right),
		$$
with pairwise correlations $\rho_{1,2}$, $\rho_{1,3}$ and $\rho_{2,3}$.

		For the normal distribution we know that the correlation of $X_1$ and $X_2$ given $X_3$ can be calculated as following
		\begin{equation*}
			\rho_{1,2 | 3} := \rho \left( X_1 | X_3, X_2 | X_3 \right) = \frac{\rho_{1,2} - \rho_{1,3} \rho_{2,3}}{\sqrt{1-\rho_{1,3}^2}\sqrt{1-\rho_{2,3}^2}}.
\label{eq:condcorr}		
		\end{equation*}
		
Defining $\rho_{1,3} = \rho_{2,3} > \rho_{1,2} > 0$ we have $\rho_{1,2 | 3} = (\rho_{1,2} - \rho_{1,3}^2)/(1-\rho_{1,3}^2) < \rho_{1,2}$, since $\rho_{1,2} \leq 1$, and $\rho_{1,2 | 3} > 0$ because of the positive-definiteness of the correlation matrix. Hence, if we fit the dependence for the two pairs with higher correlation first (assumption $\rho_{1,3} = \rho_{2,3} > \rho_{1,2} > 0$) the remaining correlation of $X_1$ and $X_2$ becomes smaller given $X_3$.
		
This is a desirable feature especially for datasets with a large number of variables, because we can truncate the R-vine specification and assume independence for the $k$ last trees to reduce the number of parameters needed. For more information on this see Section \ref{sec:applic} and \citet{BrechmannCzadoAas2010}.
\end{itemize}

We use Kendall's tau as a measure of dependence, since it measures dependence independently of the assumed distribution and hence, is especially useful when combining different (non-Gaussian) copula families. However the described method works in the same way for every other measure of dependence (see \citet[Chapter 3]{Brechmann2010} for an extensive discussion).

\citet{truncation} proposes another method to generate R-vines. She builds the trees the other way around, starting with the last tree. By this method she tries to generate an R-vine with the lowest dependencies in the top trees. This method depends on the partial correlations which contradicts the fact that we want to use other, non-Gaussian copulae. Partial correlations are used, since they can be calculated without knowing the exact R-vine structure of the first trees.

\begin{algorithm}[t]
\caption{Sequential method to select an R-vine model based on Kendall's tau.}
\label{alg:SequentialMethod}
\begin{algorithmic}[1]
	\REQUIRE Data $(x_{\ell 1}, \ldots x_{\ell n}),\ \ell=1,...,N$ (realizations of i.i.d. random vectors).
	\ENSURE R-vine copula specification, i.e., $\mathcal{V}$, $B$.
	\STATE Calculate the empirical Kendall's tau $\hat{\tau}_{j,k}$ for all possible variable pairs $\{j,k\},1\leq j<k\leq n$.
	\STATE Select the spanning tree that maximizes the sum of absolute empirical Kendall's taus, i.e.,
\begin{equation*}
\max \sum_{e=\{j,k\}\text{ in spanning tree}} |\hat{\tau}_{j,k}|.
\end{equation*} 
	\STATE For each edge $\{j,k\}$ in the selected spanning tree, select a copula and estimate the corresponding parameter(s). Then transform $\widehat{F}_{j|k}(x_{\ell j}|x_{\ell k})$ and $\widehat{F}_{k|j}(x_{\ell k}|x_{\ell j})$, $\ell=1,...,N,$ using the fitted copula $\widehat{C}_{jk}$ (see \eqref{eq:hfunc}).
	\FOR[Iteration over the trees]{$i = 2, \ldots, n-1$}
	\STATE Calculate the empirical Kendall's tau $\hat{\tau}_{j,k|D}$ for all conditional variable pairs $\{j,k|D\}$ that can be part of tree $T_i$, i.e., all edges fulfilling the proximity condition (see Definition \ref{Def:Vine}).
	\STATE Among these edges, select the spanning tree that maximizes the sum of absolute empirical Kendall's taus, i.e.,
\begin{equation*}
\max \sum_{e=\{j,k|D\}\text{ in spanning tree}} |\hat{\tau}_{j,k|D}|.
\end{equation*} 
	\STATE For each edge $\{j,k|D\}$ in the selected spanning tree, select a conditional copula and estimate the corresponding parameter(s). Then transform $\widehat{F}_{j|k\cup D}(x_{\ell j}|x_{\ell k},\boldsymbol{x}_{\ell D})$ and $\widehat{F}_{k|j\cup D}(x_{\ell k}|x_{\ell j},\boldsymbol{x}_{\ell D})$, $\ell=1,...,N,$ using the fitted copula $\widehat{C}_{jk|D}$ (see \eqref{eq:hfunc}).
	\ENDFOR
\end{algorithmic}
\end{algorithm}

Our method is summarized in Algorithm \ref{alg:SequentialMethod}. To select the tree that maximizes the sum of absolute empirical Kendall's taus (Steps 2 and 6) we use a maximum spanning tree (MST) algorithm such as the Algorithm of Prim \cite[Section 23.2]{IntroToAlgo}. Typically such algorithms are described in a way to find a \textit{minimal} spanning tree. But the algorithms work in both ways. Also note that in Steps 2 and 6 we are looking for a tree. We could look for a star or a path instead, to obtain a C- or a D-vine structure, respectively.
Note that for a D-vine a Hamiltonian path has to found which corresponds to solving a Traveling Salesman Problem. This is however NP-equivalent and therefore rather inefficient to find a solution for, especially in higher dimensions.

Notice that an MST algorithm does not depend on the the actual values of the edges, instead it only uses their rank. Therefore, the algorithm leads to the same results if we transform the edge values by a monotone increasing function. Hence, in our field of application, where we want to find a tree with maximal values of taus we would get the same tree even if we took other weights like squared taus or another monotone increasing transformation.

How to select a copula, i.e., Steps 3 and 7 of Algorithm \ref{alg:SequentialMethod} is explained in more detail in Section \ref{sec:CopulaSelection}. A proof that this algorithm creates an R-vine, i.e., that we always find a tree in Steps 2 and 6 and further explanations are given in the following.

An MST algorithm always leads to a tree when the input graph is connected. Therefore, we need to check this assumption to verify our method.

This is obviously true for $T_1$, since we start with a complete graph. When conducting the $i$-th step, we know that  $T_{i-1}$ is a tree. The node set of tree $T_i$ is then given by $N_i = E_{i-1}$. Let $E'_i$ be the set of all possible edges in $T_i$ (see Step 5 of Algorithm \ref{alg:SequentialMethod}). This edge set is defined by
\begin{equation}
	\label{eq:Estrich}
	E_i' = \{ \{a,b \} \in N_{i}^2 | \# (a \cap b) = 1\}.
\end{equation}
The requirement $\# (a \cap b) = 1$ ensures the proximity condition of an R-vine. To show that $(N_i, E'_i)$ is connected recall that connected means there is a path from every single node to every other node.
Let $a, b \in N_i$ be arbitrary nodes. Further, let $n_1,n_2 \in N_{i-1}$ be two nodes from the previous tree with $n_1 \in a\ \text{and}\ n_2 \in b$. Since $n_1$ and $n_2$ are nodes of a tree, there is a path in $T_{i-1}$ from $n_1$ to $n_2$, $n_1 \in e_1 \rightarrow \ldots \rightarrow e_l \ni n_2$, $e_1, \ldots, e_l \in E_{i-1}$, $l \geq 1$. We know that $n_1 \in a$ and $n_1 \in e_1$. Without loss of generality we can assume that $a = e_1$.
Otherwise, if $e_1 \not = a$, we can extend the path
\begin{eqnarray*}
	e_{l+1} &=& e_l \\
	&\vdots& \notag \\
	e_2 &=& e_1 \\
	e_1 &=& a \\
	l &=& l+1.
\end{eqnarray*}
Similarly we can assume $b = e_l$. Since $e_1, \ldots, e_l$ induce a path, we know that $\#(e_i \cap e_{i+1}) = 1$ for all $i = 1, \ldots, l-1$. Hence  $\{e_i , e_{i+1} \} \in E'_{i}$ for all $i = 1, \ldots, l-1$. Thus, we know that there is a path from $e_1 = a$ to $e_l = b$ and $(N_i, E'_i)$ is a connected graph. Table \ref{example:SequentialMethod} shows a concrete example of this idea.

\begin{table}[t]
\centering
\begin{tabular}{cp{0.4\linewidth}p{0.55\linewidth}}
$i$&Graph&Description\\
\hline
\hline
1&
$$
\psmatrix[colsep=1cm,rowsep=0.7cm,mnode= circle]
&1\\
2&&3\\
4&&5
\ncline{1,2}{2,1}
\ncline{1,2}{2,3}
\ncline{1,2}{3,1}
\ncline[linestyle=dashed]{1,2}{3,3}
\ncline[linestyle=dashed]{2,1}{2,3}
\ncline[linestyle=dashed]{2,1}{3,1}
\ncline[linestyle=dashed]{2,1}{3,3}
\ncline[linestyle=dashed]{2,3}{3,1}
\ncline[linestyle=dashed]{2,3}{3,3}
\ncline{3,1}{3,3}
\endpsmatrix
$$
&
\begin{flushleft}
Assume that we have 5 variables $N_1 = \{1,2,3,4,5\}$. The first graph is always a complete graph, where we can connect every node with every other node. Let us assume the Algorithm of Prim selects the solid edges. The concrete edge values (Kendall's taus) are not of interest in this example. 
\end{flushleft}
\\
\hline
2&
$$
\psmatrix[colsep=0.8cm,rowsep=0cm,mnode= oval]
1,2\\
&1,4&4,5\\
1,3
\ncline{1,1}{3,1}
\ncline[linestyle=dashed]{1,1}{2,2}
\ncline{3,1}{2,2}
\ncline{2,2}{2,3}
\endpsmatrix
$$
&
\begin{flushleft}
All edges from the previous step are now nodes. An edge is drawn whenever the nodes share a common node in the previous tree (dashed and solid). We see that the graph is connected and select the tree indicated by the solid edges.  
\end{flushleft}
\\
\hline
3&
$$
\psmatrix[colsep=0.4cm,rowsep=0cm,mnode=oval]
2,3|1&3,4|1&1,5|4
\ncline{1,1}{1,2}
\ncline{1,2}{1,3}
\endpsmatrix
$$
&
\begin{flushleft}
There are no options in this step. We need all edges to form a tree. Note, as soon as a graph has a D-vine structure, there are no more options in the following trees because they it uniquely determines all following conditioned and conditioning sets.
\end{flushleft}
\\
\hline
\end{tabular}
\caption{Exemplification of the model selection Algorithm \ref{alg:SequentialMethod}.}
\label{example:SequentialMethod}
\end{table}

Finally, we give some more insight on how to calculate the empirical Kendall's taus and select copula families. Define $E'_i$ like it was done in \eqref{eq:Estrich}. For all $e \in E'_i$ we have to calculate the value of Kendall's tau, and for some of them (those selected in the MST) we need to fit a copula based on two conditioned variables. If $e \in E'_i$, $e =\{a,b\}$ connects variables $x_{C_{e,a}}$ with $x_{C_{e,b}}$ given the variables $\boldsymbol{x}_{D_e}$, we hence need the transformed variables $F_{C_{e,a} | D_e}(x_{C_{e,a}} | \boldsymbol{x}_{D_e})$ and $F_{C_{e,b} | D_e}(x_{C_{e,b}} | \boldsymbol{x}_{D_e})$ which are obtained as described in \eqref{eq:hfunc}. For these it is then straightforward to calculate the empirical Kendall's tau and select a bivariate copula family as outlined in the following section.

\subsection{Selecting pair-copula families sequentially}\label{sec:CopulaSelection}

Besides the steps described above we need to select a copula family for every pair of variables. In the later application we take the following copula families into consideration (some properties are given in brackets):
\begin{itemize}
\item Gaussian/Normal (tail-symmetric, no tail dependence),
\item Student-t (tail-symmetric, tail dependence),
\item Gumbel (tail-asymmetric, upper tail dependence) and survival Gumbel (tail-asymmetric, lower tail dependence),
\item rotated Gumbel by 90 and 270 degrees (tail-asymmetric, no tail dependence),
\item Frank (tail-symmetric, no tail dependence).
\end{itemize}
In case of positive dependence this means that we can select among the Gaussian, Student-t, (survival) Gumbel and Frank copulae, while rotated Gumbel copulae can be used instead of Gumbel and survival Gumbel copulae when modeling negative dependence. Further, we will not use a Student-t copula if the maximum likelihood estimation leads to a degrees of freedom parameter higher than 30 because then the Student-t copula is too close to the Gaussian which can be used instead.

Given these options we still have to decide which copula fits ``best''. We do this using the AIC \citep{Akaike} which corrects the log likelihood of a copula for the number of parameters, i.e., the use of the Student-t copula is penalized compared to the other copulae, since it is the only two parameter family under consideration. Bivariate copula selection using the AIC has previously been investigated in \citet{Manner2007} and \citet[Section 5.4]{Brechmann2010} who found that it is a quite reliable criterion, in particular in comparison to alternative criteria such as copula goodness-of-fit tests. Selection proceeds by computing the AIC's for each possible family and then choosing the copula with smallest AIC. We will also include the independence copula in the selection by performing a preliminary independence test based on Kendall's tau as described in \citet{EverythingCopulaModeling}. If this test indicates independence, no further steps are taken and the independence copula is chosen.

Given the wide range of bivariate copula families available the above list of copulae clearly is not complete.
For instance, we could also consider two parameter copula families such as the BB1 or BB7 with different lower and upper tail dependence.
These have previously been used as building blocks of C- and D-vine copulae by \citet{CzadoSchepsmeierMin2010} and \citet{NikoloulopoulosJoeLi2012}.
While already including copula families able to account for very different types of dependence, the above list can easily be extended by such families, which however increases the computational burden of the copula selection step.
Using appropriate diagnostic tools for asymmetry and tail dependence as in the above two references, the required computational time can however be reduced.

\section{Modeling the residual dependency among daily returns of international financial indices}\label{sec:applic}

Copula based models are very commonly used in the area of multivariate modeling of financial returns. Here first appropriate marginal time series models are fitted to each financial return series and standardized residuals are formed. The dependency among these residuals is then modeled using a multivariate copula after a transformation to marginally uniform data using either an empirical or parametric probability integral transformation. There has been empirical evidence that different asymmetric and tail dependencies are present for different pairs of variables, which cannot be captured using a multivariate Gaussian or Student-t copula with a common degree of freedom (see, amongst others, \citet[2001]{LonginSolnik1995}\nocite{LonginSolnik2001} and \citet{AngBekaert2002}). Especially D-vines have been shown to be very successful in the modeling of such dependency patterns (see \citet{Pair-copulaConstructions}, \citet{BayesionInferenceForCopulas} and \citet{Mendes2010}), but also C-vines have recently been successfully applied \citep{CzadoSchepsmeierMin2010}. \citet{Mendes2010} however suggested that there should more research on how to choose D-vines including both the choice of the order of the nodes as well as how to choose the pair-copula families. This paper is exactly answering these questions and in our application we will investigate whether R-vine copulae other than C- or D-vine and standard multivariate copulae are needed in modeling the residual dependencies among financial returns. 

For this we selected 16 international indices, including five equity, nine fixed income (bonds) and two commodity indices observed daily from 12/29/2001 until 12/14/2009 (2337 daily returns). All returns are unhedged  against currency fluctuations and quoted  in their home currency except for global indices which are stated in USD. In particular we choose the equity indices DAX, STOXX50, S\&P500, MSCI-World  and MSCI-EE, the fixed income indices IBOXX-G-3-5, IBOXX-G-7-10, IBOXX-E-1-3, IBOXX-E-5-7, IBOXX-E-10+, IBOXX-E-A, IBOXX-E-AA, IBOXX-E-AAA, IBOXX-E-BBB and the commodity indices Comm and Gold. For the bonds we selected maturities such that those of the German and the Euro bonds are disjoint, since German bonds (IBOXX-G) account for a large proportion of the Euro indices (IBOXX-E) giving rise to extremely high Pearson correlations which are also observed between consecutive maturities (see the corresponding pairs in Figure \ref{fig:DD:T1} below). More information about the selected indices can be found in Table 6.13 of \citet{Dissmann2010}. 

\begin{figure}[t]
\centering
\includegraphics[width=0.8\textwidth]{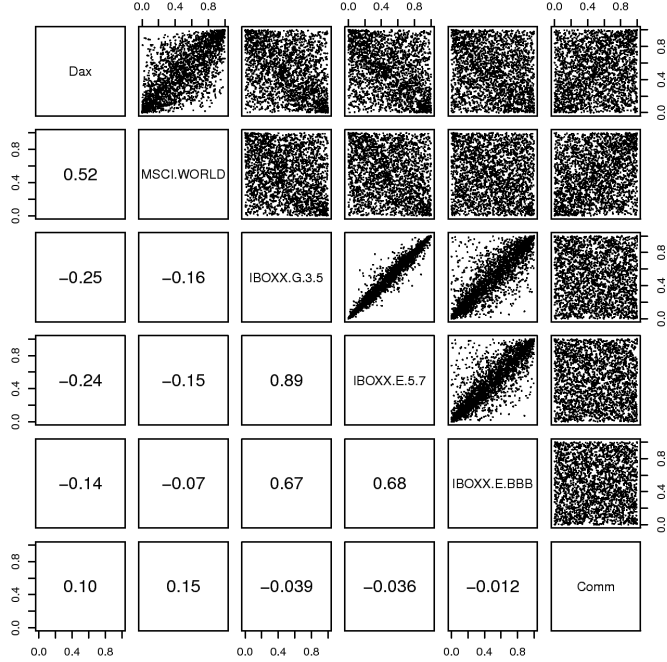}
\caption{Pairs-plots and Kendall's taus for representatives of each index group.}
\label{fig:DD:PP-ALL}
\end{figure}

For the first step we fitted univariate ARMA(1,1)-GARCH(1,1) models with Student-t innovations using maximum likelihood estimation to all equity and commodity indices and Gauss innovations for all bond indices, separate residual analyses in \citet[Section 6.3.1 and Appendix B.3]{Dissmann2010} show no volatility clusters and a good fit of the chosen innovation distribution for equity and commodity indices. For  bond indices the innovation distributions are only reasonable. Corresponding Ljung-Box tests indicate independence of the standardized residuals. Since the sample size is large and there is always some uncertainty in the innovation distribution we selected the empirical probability integral transformation to obtain marginally uniform data. The resulting pair plots of the resulting copula data (top triangular matrix) and their estimated Kendall's tau values (lower triangular matrix) for six representatives from the different indices are given in Figure \ref{fig:DD:PP-ALL} indicating different strengths and signs of pairwise dependencies. 

For model selection we want to demonstrate the superior fit of R-vines with individually chosen pair-copula families and assess the gain over R-vines with only bivariate t or with only Gauss pair-copulae as well as over standard C- and D-vines. In particular we apply the selection algorithm of Section \ref{sect:rvineselect} to select among five different R-vine classes given by

\begin{itemize}
\item {\bf mixed  R-vine}: R-vine with pair-copula terms chosen individually from seven bivariate copula types (Gauss, Student-t, Gumbel, survival Gumbel, rotated Gumbel (90 and 270 degrees), Frank).
\item {\bf mixed C-vine}: C-vine with pair-copula terms chosen individually from seven bivariate copula types (see above).
\item {\bf mixed D-vine}: D-vine with pair-copula terms chosen individually from seven bivariate copula types (see above).
\item {\bf all t R-vine}: R-vine with each pair-copula term chosen as bivariate Student-t copula. If the degrees of freedom parameter of a pair is estimated to be larger than 30, we set the copula to the Gaussian.
\item {\bf multivariate Gauss}: R-vine with each pair-copula term chosen as bivariate Gaussian copula, i.e., this corresponds to a multivariate Gaussian copula, where unconditional correlations can be obtained from conditional ones by inverting a generalized version of Equation \eqref{eq:condcorr}.
\end{itemize}

The top tree is common to all R-vines (in contrast to the C- and D-vines which are determined as maximal stars and paths as noted in Section \ref{sect:rvineselect}), since the selection of the top tree does not depend on the pair-copula choice (but only on the empirical Kendall's taus) and is given in Figure \ref{fig:DD:T1}. The structure in Figure \ref{fig:DD:T1} reflects expected relationships among the residuals of the indices. The government bond indices are grouped so that consecutive maturities are connected. Similarly corporate bond indices are aligned according to their ratings from lowest (BBB) to highest (AAA). These two groups are connected by an average representative, i.e., IBOXX-E-5-7 and IBOXX-E-AA. Since STOXX50 is a European equity index the residual dependency is highest to the predominant Euro bond index (IBOXX-G-3-5).  

\begin{figure}[t]
	\centering
	\includegraphics[width=\textwidth]{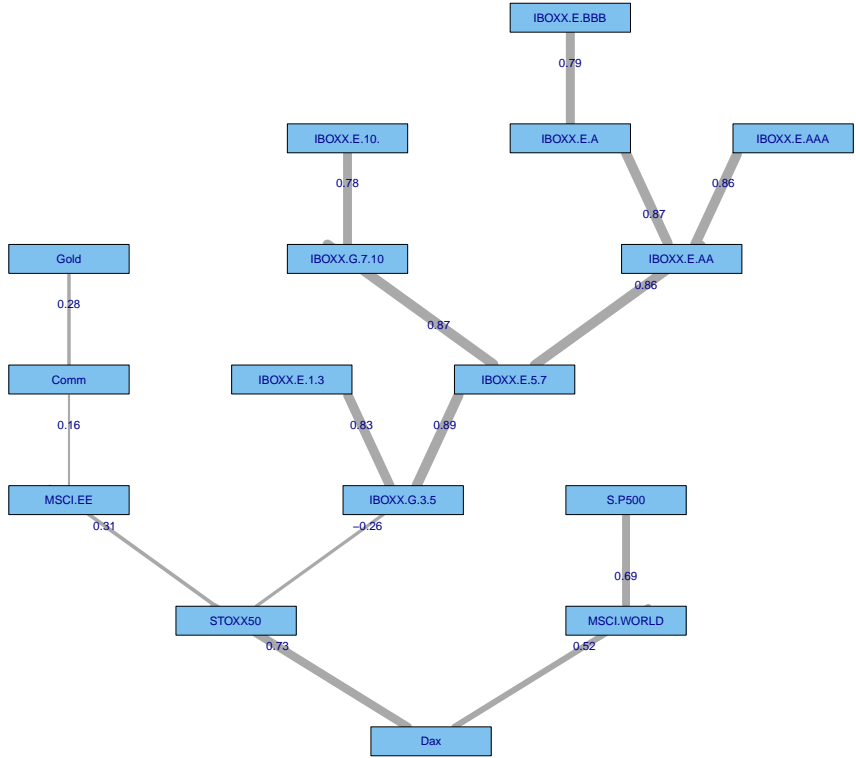}
	\caption[First tree for an R-vine for the financial indices dataset.]{$T_1$ for an R-vine from the model selection algorithm.}
	\label{fig:DD:T1}
\end{figure}

For the copula family selection of each pair-copula term the AIC is used as described in Section \ref{sec:CopulaSelection}, where pair-copula parameters are estimated by maximum likelihood estimation.
In applying the selection algorithm we also observed that empirical Kendall's tau values tend to be small for higher order trees. In this cases it might be sufficient to replace the corresponding pair-copula term by the independence copula. Therefore we also fitted an R-vine using the preliminary independence test based on Kendall's tau for each pair (``\textbf{indep. R-vine}''). If the $p$-value of the test is larger than $5\%$, then we choose the independence copula for this pair-copula term. The issue of large numbers of independence copulae in later trees is further investigated in \citet{BrechmannCzadoAas2010} who call an R-vine \textit{truncated} if all pair-copulae in higher order trees are set to bivariate independence copulae.

Applying the selection procedure to the {\bf R-vine mixed case} 16 Gauss, 51 Student-t, 4 Gumbel, 7 survival Gumbel, 12 rotated Gumbel and 30 Frank bivariate copula terms requiring 171 parameter estimates were chosen. If the choice for a pairwise independence copula is allowed, the total number of parameters was significantly reduced to 108, since 55 copula terms were replaced by an independence copula. These models correspond to the mixed/t scenario of the simulation study in \ref{app:simstudy} and hence we can assume that our models give rather adequate fits compared to the (unknown) ``true'' model.

Selection results for all models are summarized in Table \ref{tab:DD:Vuong}. It shows the log likelihood achieved for sequential estimates in the first row, while the second row gives the log likelihood after joint optimization of the chosen regular vine tree specification and copula types (see Section \ref{sect:rvineinf}). The next rows indicate the number of pair-copula types chosen and the final rows give the test statistics together with the $p$-values in parentheses of a Voung test with and without Akaike and Schwarz corrections, respectively, testing the R-vine mixed model against the alternative indicated by the respective column. This shows that the sequential log likelihood is quite close to the one obtained by joint maximization for all model classes considered. Especially the top four ranks are maintained. We also observe only small differences in the parameter estimates. The non-zero number of (survival/rotated) Gumbel pair-copula terms shows non-symmetric heavy tailed conditional dependencies present in the residual data. From the Vuong tests we see that the mixed R-vine is to be preferred over the mixed D-vine and the multivariate Gaussian copula. The difference to the all t R-vine and to the mixed C-vine is also more pronounced when using the (parsimonious) Schwarz correction, the mixed R-vine model is marginally superior in that case. The choice of Gaussian copulae for Student-t copulae with too many degrees of freedom means that the number of parameters in the all t R-vine is still close to that of the mixed R-vine. If we chose Student-t copulae for all terms, the number of parameters would be 240 and hence the influence of the corrections for the number of parameters used would be stronger. Finally, the mixed R-vine model reduced by independence pair-copula terms is preferred over the non-reduced mixed R-vine model if a Schwarz correction is used, since the reduced model has significantly less parameters to be estimated.

\begin{table}[t]
\centering
\begin{tabularx}{\textwidth}{cccccccc}
& & R-vine & R-vine & R-vine & R-vine & C-vine & D-vine\\
& & mixed & all t & all Gauss & indep. & mixed & mixed\\
\hline
\hline
\multicolumn{2}{c}{Seq. log likelihood} & 36431 & 36417 & 30445 & 36331 & 36366 & 36300\\
\multicolumn{2}{c}{Log likelihood} & 36514 & 36513 & 31784 & 36396 & 36476 & 36422 \\
\multicolumn{2}{c}{No. of parameters} & 171 & 179 & 120 & 108 & 178 & 176\\
\hline
\multirow{7}{*}{\begin{sideways}No. of copulae\end{sideways}}
& Indep. & 0 & 0 & 0 & 55 & 0 & 0\\
& Gauss & 16 & 61 & 120 & 8 & 19 & 18\\
& Student-t & 51 & 59 & 0 & 43 & 58 & 56\\
& Gumbel & 4 & 0 & 0 & 1 & 8 & 7\\
& Surv. Gumbel & 7 & 0 & 0 & 1 & 8 & 6\\
& Rot. Gumbel & 12 & 0 & 0 & 2 & 11 & 9\\
& Frank & 30 & 0 & 0 & 10 & 16 & 24\\
\hline
\multirow{6}{*}{\begin{sideways}Vuong tests\end{sideways}}
& no correction & & 0.03 & 14.59 & 6.32 & 1.00 & 3.49\\
& & & (0.97) & (0.00) & (0.00) & (0.32) & (0.00)\\ 
& Akaike corr. & & 0.49 & 14.44 & 2.92 & 1.18 & 3.68\\
& & & (0.63) & (0.00) & (0.00) & (0.24) & (0.00)\\
& Schwarz corr. & & 1.79 & 13.98 & -6.85 & 1.71 & 4.23\\
& & & (0.10) & (0.00) & (0.00) & (0.09) & (0.00)\\
\hline
\end{tabularx}
\caption{Log likelihoods, numbers of parameters and of copulae for all models as well as results of the Vuong tests (test statistics and \textit{p}-values in parantheses) comparing the R-vine model with mixed copulae to all other models. The positive values of Vuong test statistics indicate that the test favors the R-vine model over the respective alternative model (inconclusive region at the 5\%-level: $[-1.96,1.96]$).}
\label{tab:DD:Vuong}
\end{table}

Overall this example demonstrates the usefulness of R-vine copulae with individually chosen copula types for each pair-copula term. In addition the R-vine tree selection procedure gives directly economically interpretable results for this data set.

A note on the required computing time: In our implementation the sequential selection and estimation Algorithm \ref{alg:SequentialMethod} took only between 5 minutes for the reduced mixed R-vine model and 9 minutes for the mixed C-vine on a Linux cluster computer with 32 processing cores (AMD Opteron, 2.6Ghz).
In contrast the maximum likelihood estimation was computationally much more demanding.
While the computing time for the non-reduced mixed R-vine model was only 1.5 hours, it increased to about 9 hours for the all t R-vine and the mixed C- and D-vine models.

\section{Summary and discussion}

This paper provides a significant contribution towards making R-vine copulae a standard building block for copula based models. While already the introduction of C- and D-vine copulae provided flexibility in modeling dependencies, R-vine copulae provide even more modeling capabilities. Before the availability of such pair-copula constructions for multivariate copulae, the choices were rather limited. With R-vine copulae together with different choices for individual choices of copula types for each pair-copula term, the problem of too few modeling choices has shifted to the problem of too many choices to be investigated. 

In this paper we provided a general selection approach to sequentially choose the tree representation together with choosing the copula type for each copula term from a large class of bivariate copula families and estimate the corresponding parameters. The selection approach involves sequentially the use of any graph theoretic algorithm which finds a maximum spanning tree. Absolute  empirical Kendall's tau values are used as weights, but other weights are possible. In finance the use of empirical  tail dependence or other measures of joint tail behavior might be useful to investigate. 

The output of the selection procedure gives an R-vine tree structure, their corresponding pair-copula types and parameter estimates. These so-called sequential estimates can be used as starting values for determining the maximum likelihood estimates (see also \citet{Haff2010} for more details on the asymptotic behavior of these estimates). The paper also uses a matrix representation of an R-vine and provides a novel algorithm to evaluate the joint density for any arbitrary R-vine copula. The selection procedure is completely operational, it is implemented in the statistical software \textit{R} and is capable to handle medium sized dimensions of up to 20 dimensions.

As noted in Section \ref{sec:applic} it might be worthwhile to replace pair-copula terms by independence copula terms or simpler copula type choices in higher order trees. This issue has been investigated in the related work by \citet{BrechmannCzadoAas2010} who developed testing procedures to determine \textit{truncation} after a certain tree. This further balances the model flexibility with the desired parsimony of the model and opens R-vines to applications in large dimensions (see also \citet{BrechmannCzado2011}).

In future, we will also investigate the model selection problem described in Section \ref{sect:rvineselect} more closely. This includes the choice of other weights than Kendall's tau as well as the selection of C- and D-vines. In particular, the selection of the order in the first D-vine tree corresponds to a Traveling Salesman Problem and therefore is NP-equivalent. Here, tailor-made approaches for the D-vine methodology have to be considered.

\section*{Acknowledgement}

We acknowledge the helpful comments of the referees, which further improved the manuscript.
The numerical computations were performed on a Linux cluster supported by DFG grant INST 95/919-1 FUGG. 

\bibliographystyle{elsarticle-harv}
\bibliography{bib}

\appendix

\section{Simulation study}\label{app:simstudy}

In order to evaluate the approach of sequentially selecting and estimating R-vines proposed in Section \ref{sect:rvineselect}, we set up a comprehensive simulation study based on the R-vine shown in Figure \ref{fig:RVine}. In total we simulated samples of size 500, 1000 and 2000 according to twelve different scenarios, i.e., twelve different choices of pair-copula families and parameters. We repeated this 1000 times each.
The considered scenarios are:
\begin{itemize}

\item \textbf{all Gaussian, all t, all Gumbel and all Frank R-vines}: all pair-copula families are chosen as Gaussian, Student-t, Gumbel and Frank copulae, respectively. Degrees of freedom of the Student-t copula are linearly increased by 1 for pair-copula terms in higher order trees and start with 3 in the first tree.

\item \textbf{mixed R-vine}: different families for each pair-copula term.

\item \textbf{t/mixed R-vine}: Student-t copulae for pair-copulae in first two trees, mixed copulae for remaining pairs. Degrees of freedom of the Student-t copulae are also mixed.

\end{itemize}
In each of these scenarios, parameters are chosen according to two different settings of Kendall's taus (first, constant values per tree except for increased values of the ``central'' copulae $c_{2,3}$, $c_{3,6}$ and $c_{2,6|3}$, and second, mixed values; see \eqref{mat:tauconst} and \eqref{mat:taumixed}, respectively) so that we end up with twelve scenarios. While the R-vine structure matrix is given by \eqref{eq:Matrix}, corresponding matrices of Kendall's tau values as well as of copula types for the mixed and t/mixed R-vines are shown in \ref{app:matrices} below.

Having simulated from the respective true model, we sequentially select and estimate by maximum likelihood estimation an R-vine model as described above and determine the following three quantities to evaluate the adequacy of our selection and estimation approach:
\begin{itemize}

\item \textbf{general tau-difference}: we compute the mean absolute difference between pairwise empirical Kendall's taus of simulated data from the true and from the selected models. The mean over all repetitions is reported.

\item \textbf{lower} and \textbf{upper tau-difference}: similarly we compute the mean absolute difference between pairwise empirical \textit{lower} and \textit{upper exceedance Kendall's taus} which are defined for two variables $U_1$ and $U_2$ as \cite[Section 3.1.3]{Brechmann2010}
\begin{align*}
\tau^{lower}(U_1,U_2) &:= \tau(U_1,U_2|U_1\leq\delta_1,U_2\leq\delta_2)\\
\tau^{upper}(U_1,U_2) &:= \tau(U_1,U_2|U_1>1-\delta_1,U_2>1-\delta_2),
\end{align*}
and measure the strength of the joint tail behavior of $U_1$ and $U_2$. As thresholds $\delta_1$ and $\delta_2$ we choose $\delta_1=\delta_2=0.2$ as recommended by \citet{Brechmann2010}. Again the means over all repetitions are reported.

\end{itemize}
The results of the simulations are shown in Table \ref{tab:simstudy} and can be summarized as follows.

\begin{table}[t]
\begin{center}
\begin{tabular}{cccccccc}
& & \multicolumn{3}{c}{Const. Kendall's taus per tree} & \multicolumn{3}{c}{Mixed Kendall's taus} \\
\hline
 & Scenario & lower & general & upper & lower & general & upper \\ 
 & & tau-diff. & tau-diff. & tau-diff. & tau-diff. & tau-diff. & tau-diff. \\ 
\hline
\hline
\multirow{6}{*}{\begin{sideways}$N=500$\end{sideways}} & all Gauss & 0.083 & 0.015 & 0.083 & 0.087 & 0.016 & 0.087 \\ 
& all t & 0.077 & 0.019 & 0.078 & 0.080 & 0.020 & 0.082 \\ 
& all Gumbel & 0.094 & 0.018 & 0.066 & 0.098 & 0.019 & 0.073 \\ 
& all Frank & 0.101 & 0.014 & 0.100 & 0.102 & 0.015 & 0.101 \\ 
& mixed & 0.090 & 0.019 & 0.090 & 0.090 & 0.021 & 0.090 \\ 
& t/mixed & 0.079 & 0.018 & 0.080 & 0.086 & 0.018 & 0.084 \\  
\hline
\multirow{6}{*}{\begin{sideways}$N=1000$\end{sideways}} & all Gauss & 0.058 & 0.010 & 0.057 & 0.061 & 0.011 & 0.060 \\ 
& all t & 0.053 & 0.013 & 0.054 & 0.056 & 0.014 & 0.057 \\ 
& all Gumbel & 0.066 & 0.013 & 0.048 & 0.068 & 0.014 & 0.050 \\ 
& all Frank & 0.077 & 0.010 & 0.077 & 0.075 & 0.011 & 0.075 \\ 
& mixed & 0.065 & 0.016 & 0.066 & 0.065 & 0.017 & 0.065 \\ 
& t/mixed & 0.056 & 0.013 & 0.056 & 0.058 & 0.013 & 0.059 \\
\hline
\multirow{6}{*}{\begin{sideways}$N=2000$\end{sideways}} & all Gauss & 0.041 & 0.007 & 0.040 & 0.042 & 0.008 & 0.043 \\ 
& all t & 0.038 & 0.009 & 0.037 & 0.040 & 0.010 & 0.039 \\ 
& all Gumbel & 0.047 & 0.009 & 0.040 & 0.048 & 0.010 & 0.042 \\ 
& all Frank & 0.062 & 0.008 & 0.062 & 0.058 & 0.008 & 0.058 \\ 
& mixed & 0.049 & 0.013 & 0.050 & 0.048 & 0.013 & 0.048 \\ 
& t/mixed & 0.039 & 0.010 & 0.039 & 0.041 & 0.010 & 0.041 \\
\hline
\end{tabular}
\caption{Results of the simulation study. The second column indicates the respective scenario for sample sizes of $N=500$, $N=1000$ and $N=2000$. The results corresponding to the first setting of Kendall's tau values are shown in columns 3-5, while those for the second setting are displayed in columns 6-8.}
\label{tab:simstudy}
\end{center}
\end{table}

In terms of all three criteria, the performance improves with increasing sample size due to a higher estimation accuracy and the smaller simulation error. Across both settings of parameters (chosen according to Kendall's tau values), the performance is very similar and only slightly worse in the case of mixed Kendall's taus. According to the general tau-difference criterion, the (non-tail dependent) all Gaussian and all Frank R-vines are identified best. The criteria based on exceedance Kendall's taus show that the all t and the t/mixed R-vines as well as the upper tail of the all Gumbel R-vine are accurately modeled. That is our selection and estimation approach appropriately takes into account the characteristic properties of the copula models. 

Comparing the all t, the t/mixed and the mixed scenarios, it is evident that models with larger numbers of Student-t copulae (combined with mixed copulae) can be identified very well. This is in particular true when Kendall's tau values are mixed, which is typical for practical applications.

\subsection{Setting of the simulation study}\label{app:matrices}

In the following we show the matrices of Kendall's tau values for parameter choice in the above simulation study as well as the copula type matrices for the mixed and t/mixed scenarios. First, the two settings of Kendall's taus are specified as follows.
\begin{itemize}

\item Constant Kendall's taus per tree:
\begin{equation}
\tau_{const} = \begin{pmatrix}
 &  &  &  &  &  &  \\ 
0.05 &  &  &  &  &  &  \\ 
0.10 & 0.10 &  &  &  &  &  \\ 
0.15 & 0.15 & 0.15 &  &  &  &  \\ 
0.20 & 0.20 & 0.20 & 0.20 &  &  &  \\ 
0.40 & 0.40 & 0.40 & 0.40 & 0.50 &  &  \\ 
0.60 & 0.60 & 0.60 & 0.60 & 0.70 & 0.70 &  \\ 
\end{pmatrix}
\label{mat:tauconst}
\end{equation}

\item Mixed Kendall's taus:
\begin{equation}
\tau_{mixed} = \begin{pmatrix}
 \\ 
0.05 \\ 
0.10 & 0.10 \\ 
0.15 & 0.15 & 0.15 \\ 
0.20 & 0.20 & 0.20 & 0.20 \\ 
0.25 & 0.30 & 0.35 & 0.40 & 0.45 \\ 
0.50 & 0.55 & 0.60 & 0.65 & 0.70 & 0.75 & \\ 
\end{pmatrix}
\label{mat:taumixed}
\end{equation}

\end{itemize}

Using abbreviations for copula types ($N$=Gaussian, $t$=Student-t, $G$=Gumbel, $SG$=Survival Gumbel, $F$=Frank) the copula type matrices of the mixed and t/mixed scenarios are given by:
\begin{itemize}

\item mixed R-vine:
\begin{equation*}
T_{mixed} = \begin{pmatrix}
& \\ 
N & \\ 
F & N & \\ 
N & F & N & \\ 
G & SG & G & SG & \\ 
F & N & F & N & t & \\ 
SG & G & SG & G & t & t & \\
  \end{pmatrix}
\end{equation*}

\item t/mixed R-vine:
\begin{equation*}
T_{t/mixed} = \begin{pmatrix}
& \\ 
N & \\ 
F & N & \\ 
N & F & N & \\ 
G & SG & G & SG & \\ 
t & t & t & t & t & \\ 
t & t & t & t & t & t & \\
\end{pmatrix}
\end{equation*}

\end{itemize}

\end{document}